\documentclass[times, 10pt,twocolumn]{article} 
\usepackage{latex8}
\usepackage{times}
\usepackage{booktabs} 
\usepackage{url}
\usepackage{amsmath}
\usepackage{subcaption}	
\usepackage{fancyhdr}
\usepackage[export]{adjustbox}
\usepackage{graphicx}
\graphicspath{{./img/}}

\begin{document}

\title{Behavioral Malware Classification using Convolutional Recurrent Neural Networks}
\author{
        {\rm Bander Alsulami}\\
        Drexel University \\
        bma48@drexel.edu \\
	\and
        {\rm Spiros Mancoridis}\\
        Drexel University \\
        spiros@drexel.edu \\
}

\maketitle
\thispagestyle{empty}

\begin{abstract}
Behavioral malware detection aims to improve on the performance of static signature-based techniques used by anti-virus systems, which are less effective against modern polymorphic and metamorphic malware. Behavioral malware classification aims to go beyond the detection of malware by also identifying a malware's family according to a naming scheme such as the ones used by anti-virus vendors. Behavioral malware classification techniques use run-time features, such as file system or network activities, to capture the behavioral characteristic of running processes. The increasing volume of malware samples, diversity of malware families, and the variety of naming schemes given to malware samples by anti-virus vendors present challenges to behavioral malware classifiers. We describe a behavioral classifier that uses a Convolutional Recurrent Neural Network and data from Microsoft Windows Prefetch files. We demonstrate the model's improvement on the state-of-the-art using a large dataset of malware families and four major anti-virus vendor naming schemes. The model is effective in classifying malware samples that belong to common and rare malware families and can incrementally accommodate the introduction of new malware samples and families. 
\end{abstract}

\section{Introduction and Background}
Malware classification is the process of assigning a malware sample to a specific malware family. Malware within a family shares similar properties that can be used to create signatures for detection and classification. Signatures can be categorized as static or dynamic based on how they are extracted. A static signature can be based on a byte-code sequence~\cite{kolter2006learning}, binary assembly instruction~\cite{moskovitch2008unknown}, or an imported Dynamic Link Library (DLL)~\cite{schultz2001data}. Dynamic signatures can be based on file system activities~\cite{heller2003one,stolfo2005comparative}, terminal commands~\cite{wang2003one}, network communications~\cite{lee1999data,ye2001anomaly}, or function and system call sequences~\cite{schmidt2009static,jacob2008behavioral,alazab2011zero}. 

Behavioral signatures have become a useful complement to static signatures, which can be obfuscated easily and automatically~\cite{venable2007vilo}. For example, polymorphic and metamorphic malware mutate their appearance and structure without affecting their behavior~\cite{szor2005art,egele2012survey}. Behavioral features capture run-time information such as file system activities, memory allocations, network communications, and system calls during the execution of a program. Such features make behavioral malware classifiers more resilient to static obfuscation methods.

Each anti-virus vendor has a unique labeling format for malware families. The format often includes the target platform (e.g., Windows, Linux) the malware category (e.g., trojan, worm, ransomware), and an arbitrary character that describes the generation. For example, a malware sample that belongs to the ransomware family~\textit{Cerber} is labeled \textit{Ransom:Win32/Cerber.a} according to the naming scheme in the Microsoft Windows Defender anti-virus system. Such naming schemes are used to simplify the classification of malware samples, track their evolution, and associate their effective counter-response. The performance of behavioral classification models depends on the ground truth labels assigned by the various anti-virus naming schemes at training. Unfortunately, the naming schemes are inconsistent across anti-virus vendors~\cite{raiu2002virus}, which complicates the training and evaluation process. This work describes a new malware classification model that performs consistently better than other models described in previous work using various anti-virus ground truth labeling schemes.

This paper presents our contributions to behavioral malware classification using information gathered from Microsoft Windows Prefetch files. We demonstrate that our technique achieves a high classification score on common malware families for a large number of samples. We measure the generalization of our malware classification model on four different anti-virus scan engines. We demonstrate the robustness of our model on rare malware families with small sample sizes. We also evaluate the ability of our model to include the correct malware family in its top predictions. Finally, we present our model's capacity to learn the behavior of newly discovered malware samples and families. 

The paper is organized as follows: Section~\ref{sec:prefetch} explains Microsoft Windows Prefetch files, which are used as dynamic features in our model. Section~\ref{sec:related} describes previous related work. Section~\ref{sec:model} describes the architecture of our behavioral malware classification model. Section~\ref{sec:expr} explains how the dataset used in the experiment was created from the ground truth labelled data. Section~\ref{sec:eval} evaluates our model against previous work on behavioral malware classification. Finally, Section~\ref{sec:cons} outlines our conclusions and future work. 

\section{Related Work} \label{sec:related}
Behavioral malware classification has been researched extensively to mitigate the shortcomings of static malware classification. Malware that use advanced obfuscation techniques, such as polymorphism and metamorphism, are a challenge for detection and classification using static analysis techniques~\cite{filiol2006malware,marpaung2012survey}. Researchers introduced new dynamic features to profile the behavior of malware samples. They extract the program control graphs \cite{kruegel2005polymorphic} and measure the similarity between malware within the same family. The work described in \cite{rieck2011automatic,canzanese2015run,jacob2008behavioral} used sequences of function/system calls to model the behavior of malware and applied machine learning techniques to group malware with similar behavior into a common family.  

The disparity between anti-virus vendors' naming schemes affect the performance of behavioral malware classifiers \cite{bailey2007automated,canto2008large,kantchelian2015better}. A common solution is to cluster malware based on their observed behavior using unsupervised machine learning~\cite{bailey2007automated}. However, malware samples that are difficult to cluster are often left out~\cite{li2010challenges}. A method to overcome the disparity between anti-virus scan engine labels is to cluster multiple ground truth labels into a single valid ground truth source~\cite{perdisci2012vamo}. Another solution uses a method to aggregate labels in conjunction with and supervised and unsupervised machine learning techniques to infer suitable labels~\cite{kantchelian2015better}.

Our work is distinct from previous efforts in that we build a Convolutional Recurrent Neural Network that uses new dynamic features extracted from Windows Prefetch files to classify malware. The model should outperform previous work using any anti-virus labelling scheme, should perform consistenly regardless of the ground truth labels, and should be able to classify malware into both common and rare malware families. 

\section{Microsoft Windows Prefetch Files} \label{sec:prefetch}
Prefetch files contain a summary of the behavior of Windows applications. The Windows operating system uses Prefetch files to speed up the booting process and launch time of Windows programs. The Windows Cache Manager (WCM) monitors the first two minutes of the booting process and another sixty seconds after all systems services are loaded. Similarly, WCM continues to monitor the application running for ten seconds. The prefetching process analyzes the usage patterns of Windows applications while they load their dependency files such as dynamic link libraries, configuration files, and executable binary files. WCM stores the information for each application in files with a \textit{.PF} extension inside the system directory named \textit{Prefetch}.

Prefetch files store relevant information about the behaviors of the application, which can be used for memory security forensics, system resources auditing, and \textit{Rootkit} detection \cite{blunden2012rootkit,molina2008timely,malin2011malware}. Many malicious activities can leave distinguishable traces in Prefetch files \cite{malin2011malware,molina2008timely}. Even fileless malware, which are memory resident malicious programs, can leave residual trails in Prefetch files after deleting their presence from the file system \cite{dove2016fileless,benj2015fileless,Candid2016symantec}. \textit{Poweliks} is one of the first fileless malware samples that can infect a computer with Ransomware~\cite{benj2015fileless}. The malware employs evasive techniques to avoid detection from traditional anti-virus software. 

\begin{figure}[htb]
    \centering
    \includegraphics[width=0.5\textwidth,right]{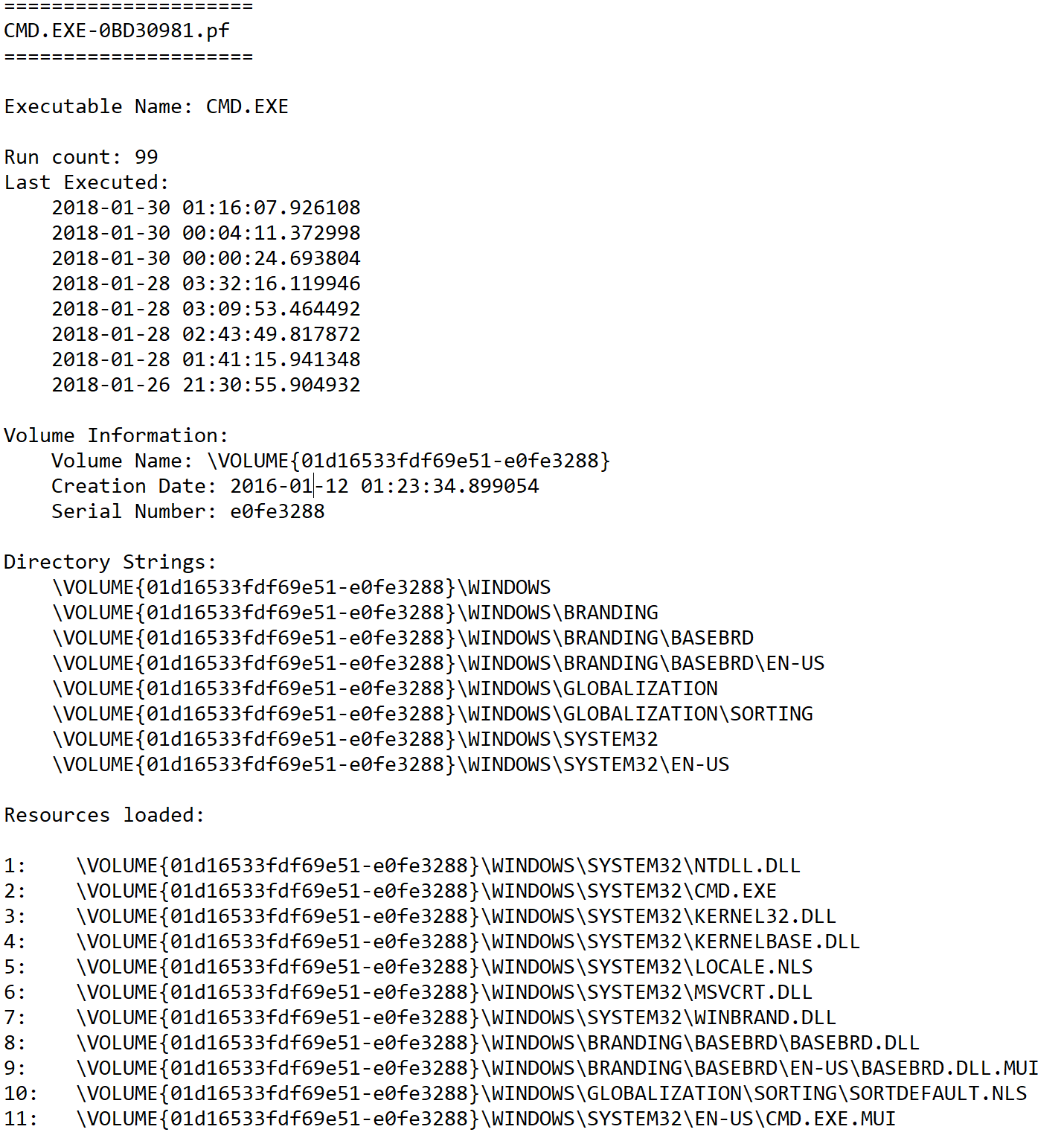}
  \caption{Example of a Prefetch file for the \textit{CMD.EXE} program}
  \label{fig:prefetch}
\end{figure}

Figure~\ref{fig:prefetch} shows an example Prefetch file for the \textit{CMD.EXE} program. The first section has runtime information such as the last-execution timestamp. The second section contains storage information. The third section lists the directories accessed by the program. The final section lists the resource files loaded by the program. The exact format of Prefetch files may vary on different versions of Windows, but the general structure is consistent across all versions. In our model, we only use the list of loaded files from the final section of each Prefetch file. 

\section{Malware Classification Model} \label{sec:model}
Our model classifies malware into families using information gathered from Prefetch files stored in the Windows Prefetch folder. We use Convolutional Recurrent Neural Networks to implement the components of our classifier. This section describes the architecture of the model and the training process used to create the model.

\subsection{Model Architecture}
\begin{figure*}[htb]
    \centering
    \includegraphics[width=0.95\textwidth,right]{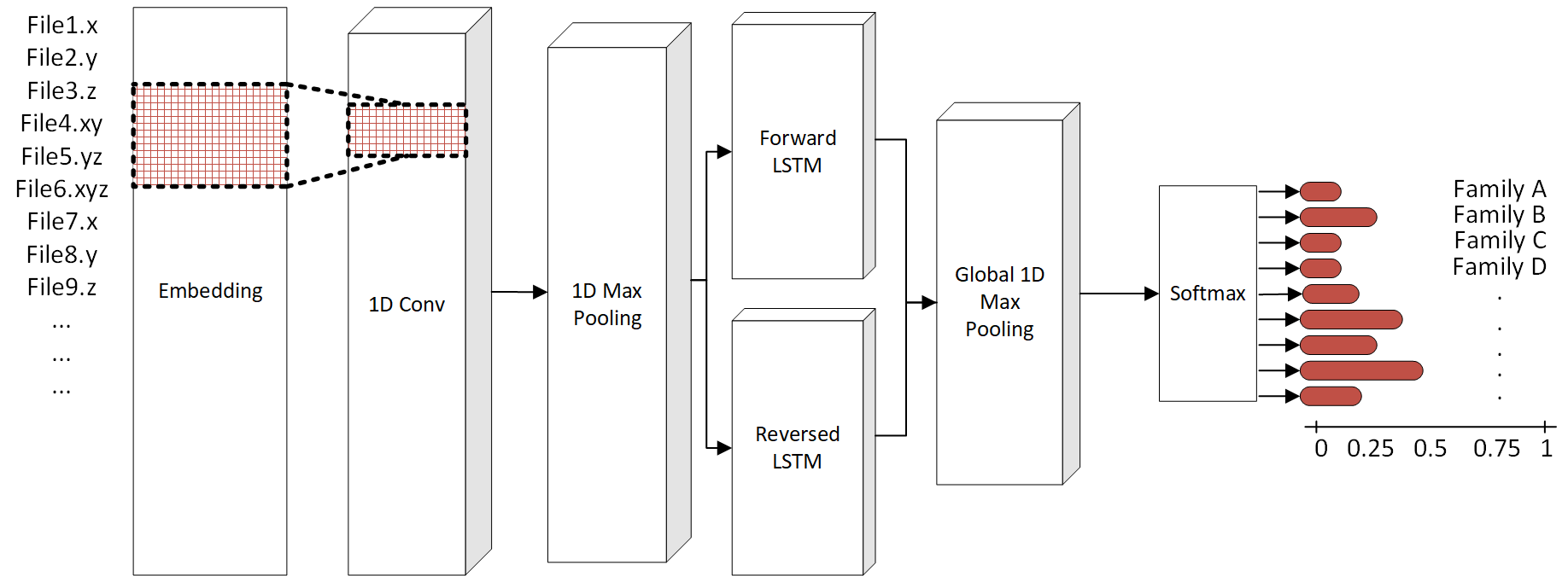}
  \caption{1D-Conv-BiLSTM model architecture}
  \label{fig:model}
\end{figure*}

Figure~\ref{fig:model} shows the general architecture of our behavioral malware classifier. The first layer is the embedding layer. This layer receives a sequence of resource file names and maps them to embedding vectors of arbitrary sizes. The number of embedding vectors represents the size of the vocabulary of the model. Each file name corresponds to a unique embedding vector. Embedding vectors generally improve the performance of large neural networks for complex learning problems~\cite{pennington2014glove}.
 
The second layer is a convolutional layer. The layer applies a one dimensional (1D) sequential filter of a particular size. The layer, then, slides the filter over the entire list to extract adjacent file names. This helps the model learn the local relation between embedding vectors. 1D convolutional layers have been used successfully in sequence classification and text classification~\cite{kim2014convolutional} problems. 

The third layer is Max Pooling. This layer reduces the size of the data from the previous layer. It is designed to improve the computational performance and the accuracy of our model and its respective training process. We use the \textit{maximum} function to select the important representation out of the data. 

The fourth layer is Bidirectional LSTM. Bidirectional LSTM (BiLSTM) is an  architecture of recurrent neural networks~\cite{goodfellow2016deep}. Recurrent neural networks learn the long-term dependency between the embedding vectors. In our context, they model the relationship between the resources file names loaded in each Prefetch file. The bidirectional structure consists of a forward and reversed LSTM, which is a structure that has been successful in NLP and sequence classification problems~\cite{wu2016google,graves2005framewise}. 

The fifth layer is Global Max Pooling. This layer propagates only relevant information from the sequence of outputs of BiLSTM. It reduces the size of the output of the BiLSTM layer. 

The sixth, and final, layer is Softmax. This layer outputs the probability that a malware sample belongs to a specific malware family. 

To improve the generalization of our model, we apply different regularization techniques. First, we apply dropout between our model layers. Dropout is a commonly used technique in training large neural networks to reduce overfitting~\cite{srivastava2014dropout}.  Dropout has shown to improve the training and classification performance of large neural networks. The goal is to learn hidden patterns without merely memorizing the training samples in the training data. This improves the robustness of the model on unseen (i.e., zero-day) malware samples. 

\section{Experimental Setup}  \label{sec:expr}
This section described how the dataset and the ground truth labeling used in our experiment was created.

\subsection{Dataset Collection}
We successfully executed around 100,000 malware samples obtained from the public malware repository VirusShare\footnote{VirusShare\footnote, http://www.virusshare.com}. Malware samples were deployed on freshly installed Windows 7 executing on a virtual machine. After each Prefetch file is collected, the virtual machine is reset to a clean (non-infected) state. In order for Windows to generate a Prefetch file for malware sample, the sample needs to be executed. Once the sample is loaded, Windows generates a Prefetch file automatically. This simplifies the task of extracting the Prefetch files for malicious programs. Our experiments only included malware samples that produced Prefetch files and were identified by major anti-virus engines, such as Kaspersky, EsetNod32, Microsoft, and McAfee.

\begin{table}[]
\centering
\begin{tabular}{|p{0.3\linewidth}|p{0.1\linewidth}|p{0.6\linewidth}|}    \hline
\begin{tabular}[c]{@{}l@{}}\textbf{Type}\end{tabular} & \textbf{Size} & \textbf{Malware Family Samples} \\ \hline
\textbf{Adware}                                                                                   & 0.79\%                         & MultiPlug, SoftPulse, DomaIQ                     \\ \hline
\textbf{Backdoor}                                                                                 & 2.25\%                         & Advml, Fynloski, Cycbot, Hlux                    \\ \hline
\textbf{Trojan}                                                                                   & 89.18\%                        & AntiFW, Buzus, Invader, Kovter                   \\ \hline
\textbf{Virus}                                                                                    & 1.44\%                         & Lamer, Parite, Nimnul, Virut                     \\ \hline
\textbf{Worm}                                                                                     & 4.28\%                         & AutoIt, Socks, VBNA, Generic                     \\ \hline
\textbf{Ransomware}                                                                                   & 2.07\%                         & Xorist, Zerber, Blocker, Bitman                  \\ \hline
\end{tabular}
\caption{Malware types, size, and examples of malware families according to the Kaspersky, EsetNod32, Microsoft, and McAfee}
\label{table:malwarefamily}
\end{table}

\subsection{Ground Truth Labeling}
Ground truth labels for malware were obtained through an online third-party virus scanning service called VirusTotal\footnote{VirusTotal, http://www.virustotal.com}. Given an MD5, SHA1 or SHA256 of a malware file, VirusTotal provides the detection information for popular anti-virus engines. This information also includes meta-data such as target platforms, malware types, and malware families for each anti-virus scan engine. Table~\ref{table:malwarefamily} illustrates malware types, sample size, examples of malware families according to EsetNod32, Kaspersky, Microsoft, and MacAfee.

\section{Evaluation} \label{sec:eval}
This section describes the experimental evaluation of our model against a model from previous work.

\subsection{Performance Measurements}
The classification accuracy of our classification model is measured by the F1 score, F1 demonstrates the trade-off between Recall and Precision and combines them into a single metric range from 0.0 to 1.0. Recall is the fraction of a number of retrieved examples over the number of all the relevant examples. Precision is the fraction of the number of relevant examples over the number of all retrieved ones. The F1 score formula is:
\begin{align*}
F1 = 2 * \frac{Precision*Recall}{Precision+Recall}
\end{align*}

A classifier is superior when its F1 score is higher. We choose the F1 score because it is less prone to unbalanced classes in training data~\cite{chawla2005data}. Malware training datasets often contain unbalanced samples for different malware families. The ratio between malware family sizes sometimes varies 1:100. Table~\ref{table:classification} shows malware type, size of malware type, and a few examples of malware families. 

\begin{table*}[th]
\centering
\begin{tabular}{c|c|c|c|c|c}
\cline{2-5}
                                     & \multicolumn{4}{c|}{Anti-virus label (\# of malware families)}    &                                     \\ \hline
\multicolumn{1}{|c|}{}               & Kaspersky (50)  & EsetNod32 (53)  & Microsoft (38)  & McAfee (55)     & \multicolumn{1}{c|}{F1 mean}        \\ \hline
\multicolumn{1}{|c|}{1D-Conv-BiLSTM} & \textbf{0.734} & \textbf{0.854} & \textbf{0.754} & \textbf{0.765} & \multicolumn{1}{c|}{\textbf{0.777}} \\ \hline
\multicolumn{1}{|c|}{LR 2-grams}     & 0.711          & 0.821          & 0.734          & 0.756          & \multicolumn{1}{c|}{0.756}          \\ \hline
\multicolumn{1}{|c|}{LR 3-grams}     & 0.718          & 0.822          & 0.726          & 0.756          & \multicolumn{1}{c|}{0.756}          \\ \hline
\multicolumn{1}{|c|}{RF 2-grams}     & 0.702          & 0.792          & 0.731          & 0.755          & \multicolumn{1}{c|}{0.745}          \\ \hline
\multicolumn{1}{|c|}{RF 3-grams}     & 0.671          & 0.699          & 0.72           & 0.724          & \multicolumn{1}{c|}{0.704}          \\ \hline
\end{tabular}
\caption{F1 score for 1D-Conv-BiLSTM, LR (2,3)-grams, and RF (2,3)-grams models using Kaspersky, EsetNod32, Microsoft, and McAfee labelings.}
\label{table:classification}
\end{table*}

\subsection{Classification Performance with Common Malware Families}
We evaluate our malware classification model against the model of previous work on behavioral malware classification~\cite{canzanese2015run}. The previous work examined multiple types of feature extractions, feature selections, classification models based on large datasets extracted from sequences of OS system calls. The top performing models were Logistic regression (LR) and Random Forests (RF). LR and RF were used with n-grams feature extraction and \textit{Term Frequency-Inverse Document Frequency} (TF-IDF) feature transformation~\cite{cavnar1995using}. RF also used \textit{Singular Value Decomposition} (SVD) for feature dimensionality reduction~\cite{kim2005dimension}. 

We implemented our new model using the Keras and Tensorflow~\cite{chollet2017keras,abadi2016tensorflow} deep learning frameworks. We configured our model using the following parameters:
\begin{itemize}
\item Embedding layer: 100 hidden units 
\item 1D Convolutional layer: 250 filters, kernel size of five, one stride, and RELU activation function
\item 1D Max Pooling: pool size of four
\item Bidirectional LSTM: 250 hidden units
\item L2 regularization: 0.02
\item Dropout regularization: 0.5
\item Recurrent Dropout regularization: 0.2
\end{itemize} 
We implemented the previous work LR and RF models using Scikit-learn~\cite{pedregosa2011scikit}. We applied a grid search to select the best hyperparameters for the LR and RF models. 

We train our model using Stochastic Gradient Descent (SGD) with batch size of 32 samples and 300 epochs~\cite{zhang2004solving}. SGD is an iterative optimization algorithm commonly used in training large neural networks. SGD can operate on large training sets using one sample or a small batch of samples at a time. Thus, it is efficient for large training sets and for online training~\cite{bottou2010large}.

We use a 10-fold cross-validation with stratified sampling to create a balanced distribution of malware samples for malware families in each fold. We train the models on 9 splits of our dataset and test on a separate dataset. We repeat this experiment 10 times and take the average metric score for the final output. We include any malware families that have a minimum of 50 malware samples. 

Table~\ref{table:classification} shows the F1 score results of our experiment using four major anti-virus scan engines: Kaspersky, EsetNode32, Microsoft, and MacAfee. The results show that our model outperforms all other models using any anti-virus engine labeling. The second best are the LR models, which outperform the RF models on all anti-virus scan engines and reproduce the results described in~\cite{canzanese2015run}. It is noteworthy that the 3-gram features extraction usually provides better results than the 2-gram features in the LR models. However, the 2-gram features outperform the 3-gram features in the RF models. 

As shown, the performance of behavioral classification models depends on the anti-virus engine labelling scheme used during training. LR 3-grams show a better performance using the Kaspersky and EsetNode32 labelings, while a worse performance using the Microsoft labeling scheme. Moreover, RF 2-grams underperform all LR models except when using the Microsoft naming scheme. The inconsistency of the results leads researchers to use the anti-virus engine that produces the highest classification score. However, our model shows consistent performance across all major anti-virus engines and outperforms previous work on major anti-virus engines.

\subsection{Classification Performance with Rare Malware Families}
Rare malware families with small sample sizes represent a significant percentage of all malware families. This presents a difficulty for models to extract useful behavioral patterns due to insufficient samples during training. In this experiment, we include any malware family that has at least 10 malware samples. This presents a challenge for classification models because the number of malware families largely increases while, at the same time, the number of malware samples for each family decreases. We aim to show the robustness of our classification model when applied to rare malware families. 

Table~\ref{table:rare} shows the classification performance of our model against LR and RF models using four anti-virus labeling schemes. The table shows that our model consistently outperforms all other models despite the increased number of malware families with a low sample size. For example, on the EsetNod32 labeling scheme, our model performance decreases only -1.0\% when the number of families increases from 53 to 180 families while other models exhibit larger classification performance degradations. Specifically, our model shows the smallest decrease in the classification performance from any anti-virus labeling scheme. 

\begin{table*}[]
\centering
\begin{tabular}{c|c|c|c|c|c|c|c|c|cc}
\cline{2-9}
\multicolumn{1}{l|}{}                & \multicolumn{8}{c|}{Anti-virus label (\# of malware families)}                                                                                         & \multicolumn{1}{l}{}                & \multicolumn{1}{l}{}                  \\ \cline{2-9}
                                     & \multicolumn{2}{c|}{Kaspersky (192)} & \multicolumn{2}{c|}{EsetNod32 (180)} & \multicolumn{2}{c|}{Microsoft (137)} & \multicolumn{2}{c|}{McAfee (209)} &                                     &                                       \\ \cline{2-11} 
                                     & F1               & Diff (\%)         & F1               & Diff (\%)         & F1               & Diff (\%)         & F1              & Diff (\%)       & \multicolumn{1}{c|}{F1 mean}        & \multicolumn{1}{c|}{Diff (\%)}   \\ \hline
\multicolumn{1}{|c|}{1D-Conv-BiLSTM} & \textbf{0.647}   & \textbf{-0.088}   & \textbf{0.844}   & \textbf{-0.010}   & \textbf{0.727}   & \textbf{-0.027}   & \textbf{0.720}  & \textbf{-0.045} & \multicolumn{1}{c|}{\textbf{0.735}} & \multicolumn{1}{c|}{\textbf{-4.25\%}} \\ \hline
\multicolumn{1}{|c|}{LR 2-Grams}     & 0.586            & -0.124            & 0.790            & -0.032            & 0.656            & -0.078            & 0.652           & -0.104          & \multicolumn{1}{c|}{0.671}          & \multicolumn{1}{c|}{-8.45\%}          \\ \hline
\multicolumn{1}{|c|}{LR 3-Grams}     & 0.594            & -0.124            & 0.790            & -0.032            & 0.651            & -0.075            & 0.656           & -0.100          & \multicolumn{1}{c|}{0.673}          & \multicolumn{1}{c|}{-8.28\%}          \\ \hline
\multicolumn{1}{|c|}{RF 2-Grams}     & 0.588            & -0.114            & 0.760            & -0.031            & 0.664            & -0.067            & 0.658           & -0.097          & \multicolumn{1}{c|}{0.668}          & \multicolumn{1}{c|}{-7.73\%}          \\ \hline
\multicolumn{1}{|c|}{RF 3-Grams}     & 0.527            & -0.144            & 0.650            & -0.049            & 0.627            & -0.093            & 0.587           & -0.137          & \multicolumn{1}{c|}{0.598}          & \multicolumn{1}{c|}{-10.58\%}         \\ \hline
\end{tabular}
\caption{F1 score for 1D-Conv-BiLSTM, LR (2,3)-grams, and RF (2,3)-grams models using Kaspersky, EsetNod32, Microsoft, and McAfee labelings. Diff (\%) shows the change of the F1 scores from the previous section after adding rare malware families.}
\label{table:rare}
\end{table*}

\begin{figure}[]
    \centering
    \includegraphics[width=0.45\textwidth,center]{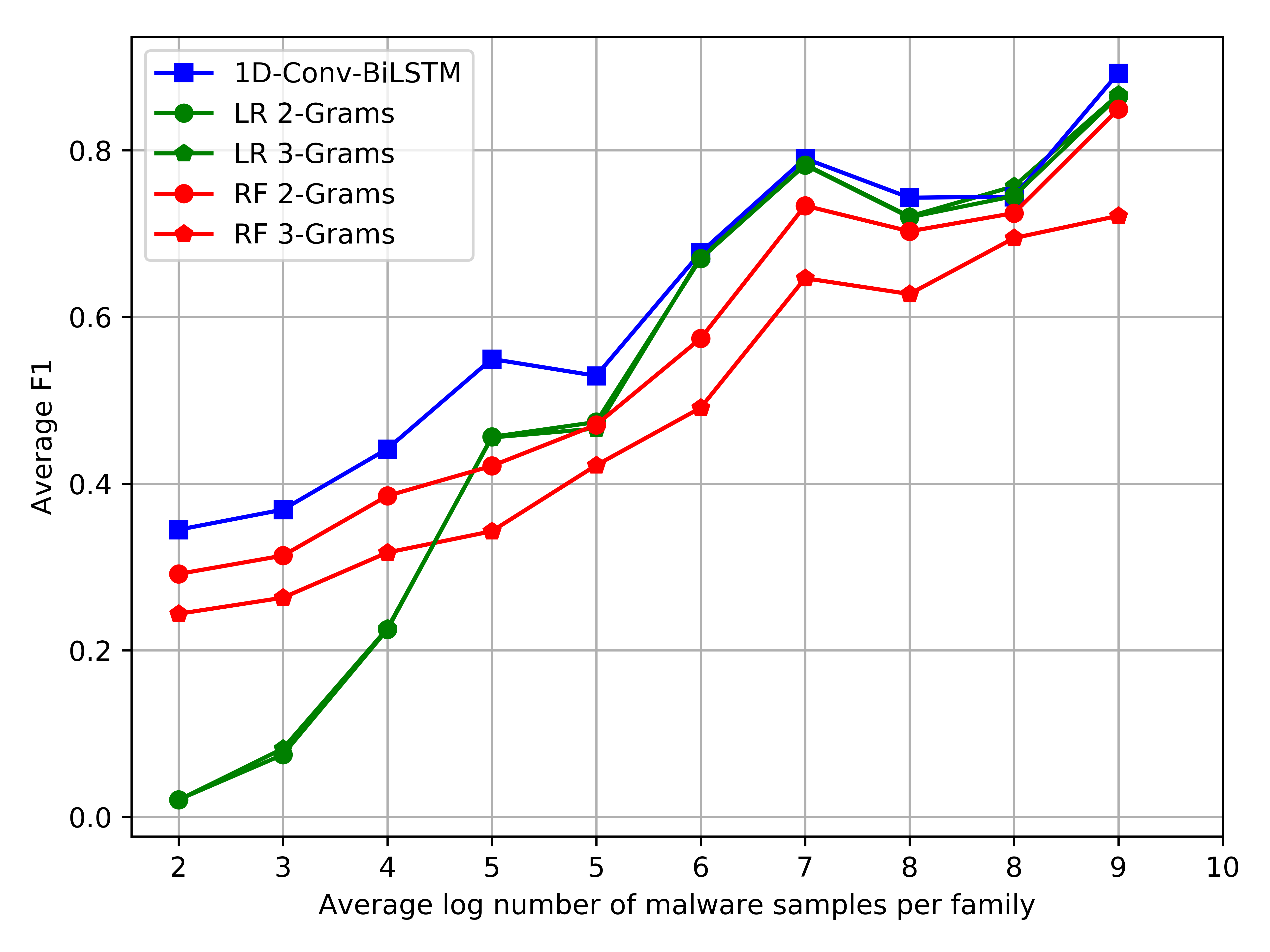} 
  \caption{Average F1 scores of the log number of malware samples per family for 1D-Conv-BiLSTM, LR 3-grams, and RF 2-grams using EstNod32 ground truth labels. }
  \label{fig:families}
\end{figure}

Figure~\ref{fig:families} shows the average F1 scores of malware families for LR 3-grams, RF 2-grams, and 1D-Conv-BiLSTM using EsetNod32 ground truth labels. We study the performance of the behavioral classification models on individual malware families to demonstrate the strength of the classification models on common and rare malware families. As shown, the LR model struggles with rare malware families. However, it outperforms the RF model when the number malware samples in a family increases. Conversely, the RF model performs reasonably on rare malware families, but it underperforms the LR models on common malware families. Ultimately, our 1D-Conv-BiLSTM model outperforms both LR and RF models on almost all common and rare malware families.

\subsection{Top Predictions Performance}
\begin{figure}[]
    \centering
    \includegraphics[width=0.45\textwidth,center]{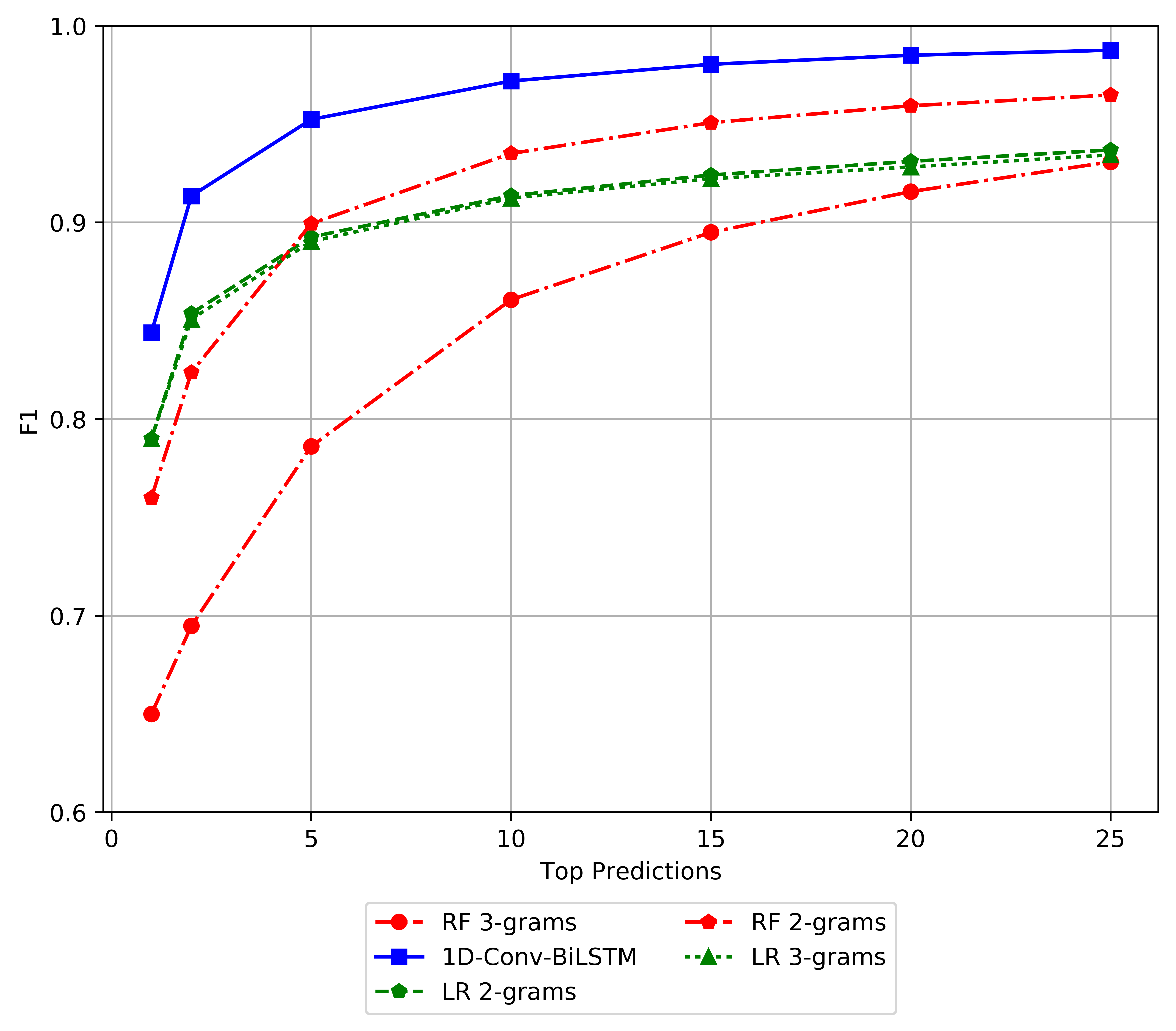} 
  \caption{The F1 scores for behavioral classification models when top k predictions are used to find the correct malware family label according to EsetNod32 ground truth labeling. }
  \label{fig:predicitions}
\end{figure}

We also evaluated the capacity of the classification models to find the correct malware family label considering their top \textit{k} predictions. That is, how the F1 score improves when the top [1,2,...,k] predictions include the correct malware family label. As shown in Figure~\ref{fig:predicitions}, 1D-Cov-BiLSTM consistently outperforms all of the other models using the top [1,2,...,25] predictions. 1D-Conv-BiLSTM achieves around 0.91, 0.95, and 0.99 F1 on the top 2, 5, and 25 predictions, respectively. This demonstrates that the correct malware family label is usually 99\% within the top 25 predictions of our model. The performance of the RF models vary between the (2,3)-grams models, while the LR models achieve similar F1 scores between (2,3)-grams models using top predictions. 

The LR (2,3)-grams models outperform the RF models up to the top 5 predictions. Then, the RF 2-grams model outperforms the LR models on the top 5 or higher predictions. The RF 3-grams model, which achieves the lowest classification performance in our experiment, matches the corresponding LR model performance when considering the top 25 predictions. This shows that RF models have a higher capacity to find the correct malware families within the top candidates. The reason might be related to the fact that a Random Forest is an ensemble of decision trees~\cite{breiman2001random}, and it is knowns that ensemble models often overcome the limitation of stand-alone classification models~\cite{wozniak2014survey}. Our model consistently outperforms the LR and RF models on the top \textit{k} predictions. 

\subsection{Classification Performance with New Malware Families}
Behavioral malware classification models need to learn the behavior of newly discovered malware continuously. This presents a challenge since the rate of malware sample discovery is high. Therefore, it is efficient, and practical, incrementally to train an existing model rather than re-train it from scratch on newly discovered samples. Incremental training provides a practical solution to assimilate new malware behavioral information into the classification models without impacting the classification performance. 

In this experiment, we evaluate our pre-trained model's ability to learn the behavior of new malware samples quickly. We train our model on all malware families that were discovered from 2010-2016. Then, we add malware families that were discovered in 2017 to the training dataset and incrementally retrain the model to create a new classification model. We aim to show that incrementally re-training an existing model is more efficient and adaptive than training a new model from scratch. 

\begin{figure}[]
    \centering
    \includegraphics[width=0.45\textwidth,center]{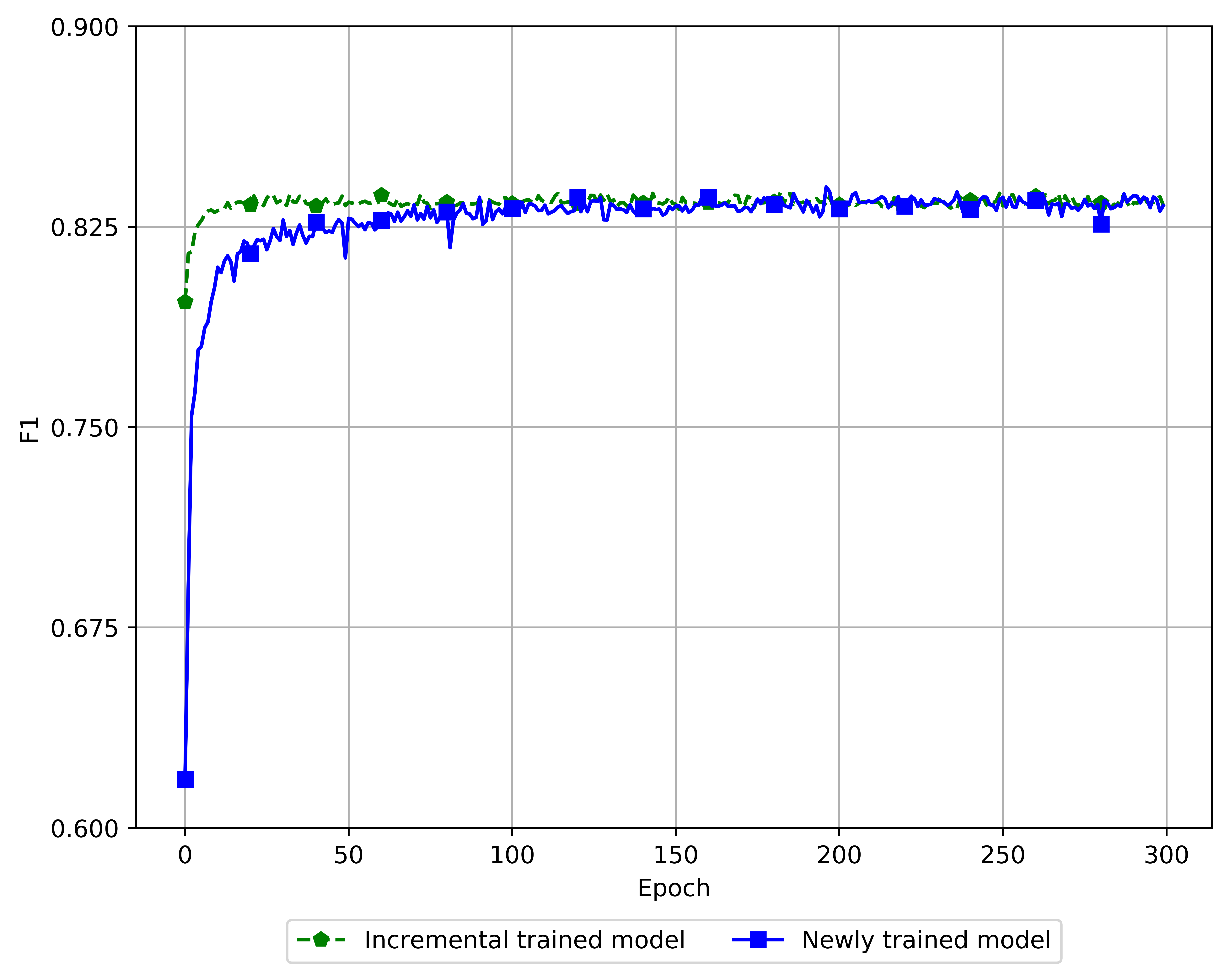} 
  \caption{The F1 scores for newly trained and incremental trained 1D-Conv-BiLSTM models on the test dataset during training. }
  \label{fig:incremental}
\end{figure}

Figure~\ref{fig:incremental} shows the classification performance of our models during training. The experiment shows that the incrementally re-trained model achieves a higher F1 at early stages during training than the newly trained model. Therefore, the training process can be shortened to reduce the overhead of training on new malware samples. Moreover, incremental re-training of our model is efficient and recommended over fully re-training the model.

\section{Conclusion} \label{sec:cons}
We introduce a new behavioral malware classification model for the Microsoft Windows platform. Our model extracts features from the Windows Prefetch files. We show the effectiveness of our classification technique on a large malware collection and ground truth labels from 4 major anti-virus vendors. 

We also evaluate our models on rare malware families with a small number of malware samples. Despite the increasing number of malware families, our model still outperforms other state-of-the-art models. Moreover, we demonstrate our model's ability to continuously learn the behavior of new malware families, which reduces the time and overhead of the training process.  

In the future, we would like to improve our ground truth labeling by combining all major scan engine labels to increase the performance and robustness of our classification model. We would also like to test our model on evolving malware families over time.

\nocite{ex1,ex2}
\bibliographystyle{latex8}
\bibliography{latex8}

\end{document}